%% file: eprint.tex
\documentclass[12pt]{article}
\usepackage{graphicx}
\usepackage[export]{adjustbox} 
\usepackage{amsmath}
\usepackage{float}
\usepackage{hyperref}
\usepackage{xspace}
\usepackage[numbers,sort&compress]{natbib}

\newcommand\pubnumber{CMS-CR-2024-005}
\newcommand\pubdate{January 9, 2024}
\newcommand{\POWHEG} {{\textsc{powheg}}\xspace}

\def\institute{Helsinki Institute of Physics, Helsinki, Finland}

\def\Title#1{\begin{center} {\Large #1 } \end{center}}
\def\Author#1{\begin{center}{ \sc #1} \end{center}}
\def\Address#1{\begin{center}{ \it #1} \end{center}}

\newcommand\pubblock{\rightline{\begin{tabular}{l} \pubnumber\\
         \pubdate  \end{tabular}}}
\newenvironment{Abstract}{\begin{quotation}  }{\end{quotation}}
\newenvironment{Presented}{\begin{quotation} \begin{center} 
             PRESENTED AT\end{center}\bigskip 
      \begin{center}\begin{large}}{\end{large}\end{center} \end{quotation}}

\input econfmacros.tex

\begin{document}
\begin{titlepage}
{\let\thefootnote\relax
\footnotetext{\hspace*{-18pt}%
Copyright 2024 CERN for the benefit of the ATLAS and CMS collaborations. Reproduction of this article or parts of it is allowed as specified in the CC-BY-4.0 license.}}
\pubblock

\vfill
\Title{Top mass measurements}
\vfill
\Author{ Mikael Myllymäki\\ on behalf of the ATLAS and CMS Collaborations}
\Address{\institute}
\vfill
\begin{Abstract}
The top quark mass measurements are based either on a direct kinematic reconstruction of the top quark decay products or on indirect measurements, where an observable sensitive to the top quark mass, such as the production cross section, is used to infer it. The ATLAS and CMS collaborations have measured the top quark mass using various methods with increasing precision. Recent measurements using 13 TeV pp collision data recorded at the LHC are presented in this review.
\end{Abstract}
\vfill
\begin{Presented}
$16^\mathrm{th}$ International Workshop on Top Quark Physics\\
(Top2023), 24--29 September, 2023
\end{Presented}
\vfill
\end{titlepage}
\def\thefootnote{\fnsymbol{footnote}}
\setcounter{footnote}{0}

\section{Introduction}
The top quark is the heaviest elementary particle in the standard model (SM)
of particle physics.
The mass of the top quark is a fundamental parameter in the SM and intriguing due to 
its connection to the stability of the electroweak (EW) vacuum \cite{stability}. 
It also serves as an input to global EW fits crucial for the SM consistency tests.
However, top quark decays are challenging to model and it is not straightforward 
to relate the reconstructed mass to a theoretically better defined mass scheme 
as discussed in Ref.~\cite{Hoang}.
Therefore, different approaches with varying analysis strategies and leading 
uncertainty sources are needed for a reliable overall view.

From the top quark mass measurements conducted by the ATLAS~\cite{Atlas} and CMS~\cite{CMS} experiments 
during Run 2 (2016-2018) with $\sqrt{s}=13$ TeV pp collision data at the LHC~\cite{LHC},
five different approaches are presented. 
In so-called direct measurements, the decay products are used to reconstruct the top 
quark, whereas indirect measurements infer the mass from a cross-section 
measurement.
The most precise direct measurement utilizing the profile likelihood method leads to 
a value of $m_{\text{t}} = 171.77 \pm 0.37$ GeV \cite{CMS-PAS-TOP-20-008}.
The most precise cross-section based measurement is reported as 
$m_{\text{t}}^\text{pole} = 170.5\pm 0.8$ GeV \cite{poleprecise}.

\section{Measurements}
\noindent
\textbf{CMS lepton + jets}\\
This measurement performed by the CMS experiment using $\text{t}\bar{\text{t}}$ events in the lepton and jets channel corresponding to an integrated luminosity of $36.3$ fb$^{-1}$ is presented in Ref.~\cite{CMS-PAS-TOP-20-008}.
One muon or electron and four jets are selected, of which two are identified as originating from b quarks using specific b tagging algorithms. 
The remaining two light-quark jets are associated to the W boson decay.

A kinematic fit based on the parton-object resolution functions is applied for selecting the best event hypothesis using the $\chi^2$ minimization.
Constraints for the W boson and top masses are used in the fit, such that  $m_\text{W}^{\text{fit}} = 80.4$ GeV and $m_\text{t}^{\text{hadronic}} = m_\text{t}^{\text{leptonic}}$.
A goodness-of-fit value P$_\text{gof}$ is derived from the $\chi^2$ value using $\text{P}_{\text{gof}} = \exp (-\chi^2/2)$, for which a default cut-off value of 0.2 is applied.

\begin{figure}[H]
\centering
\includegraphics[width=0.395\textwidth]{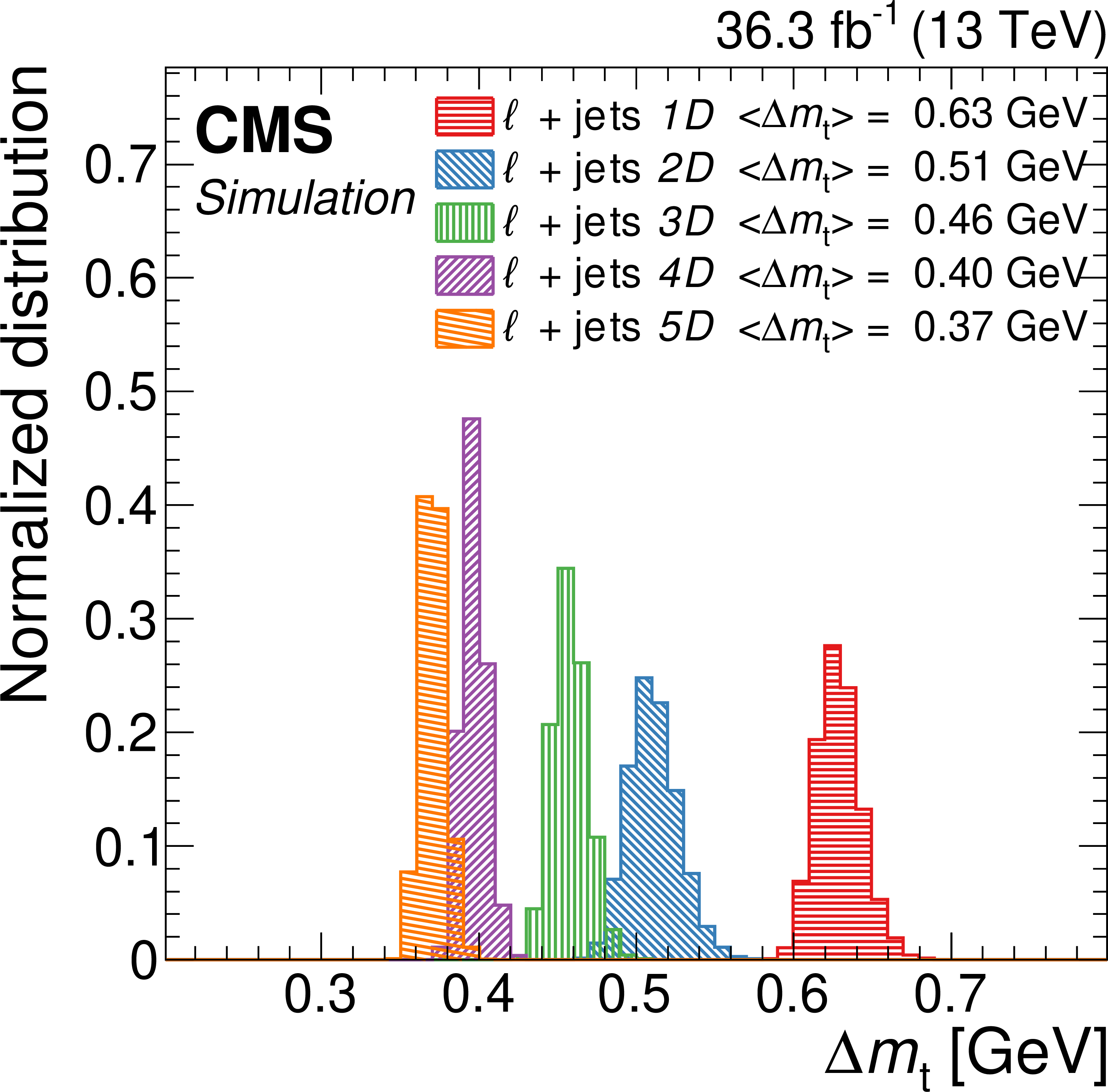}
\includegraphics[width=0.595\textwidth]{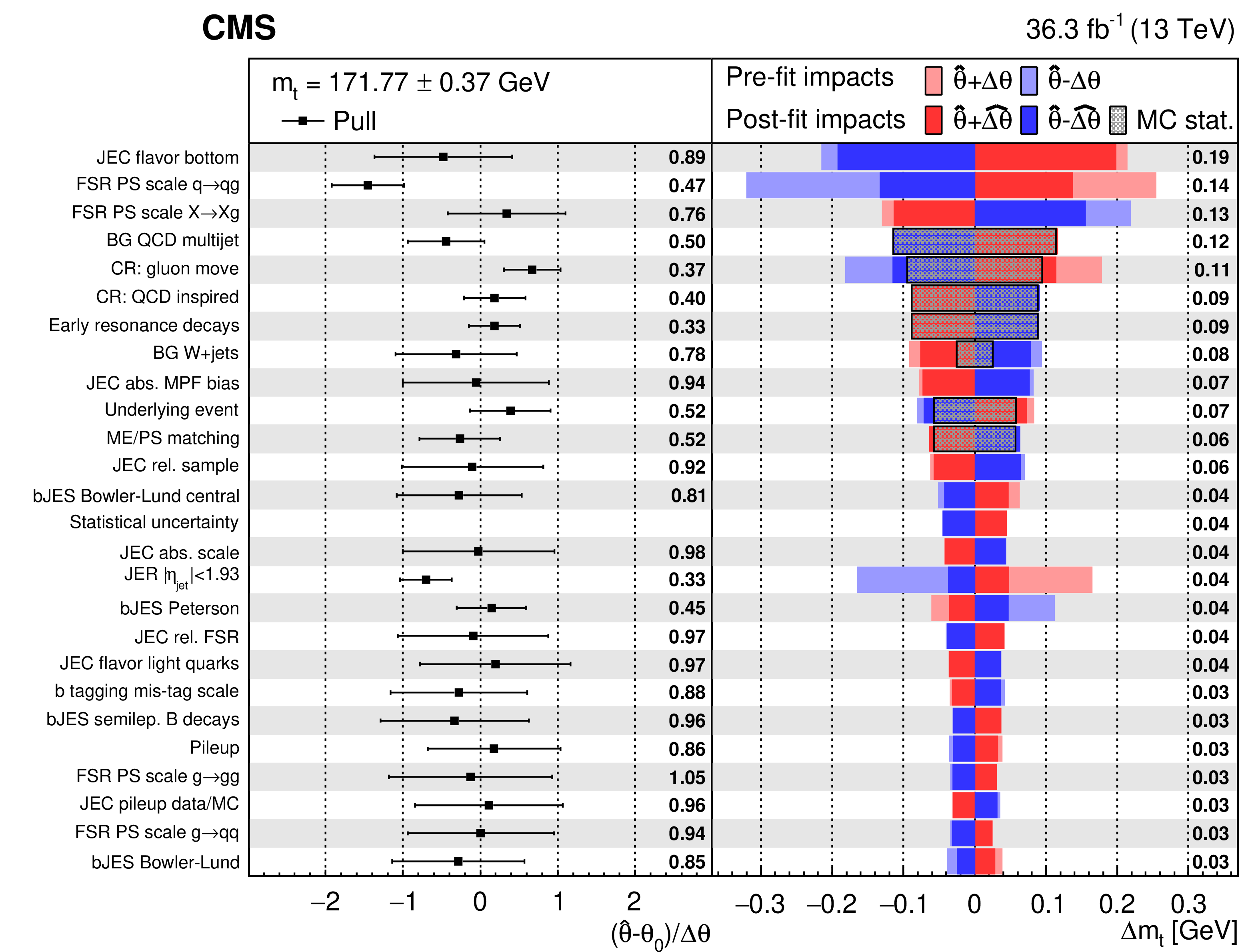}
\caption{Left: Expected total uncertainties on $m_\text{t}$ for different number of observables.
Right: Post-fit pulls and pre-fit and post-fit impacts for the most important nuisance parameters
in the lepton+jets channel. \cite{CMS-PAS-TOP-20-008}}
\label{fig:ljets}
\end{figure}

Mass is extracted using a profiled maximum-likelihood (ML) method with five observables: $m_{\text{t}}^{\text{fit}}$, $m_\text{W}^{\text{reco}}$, $R_{\text{bq}}^{\text{reco}}$, $m_{\ell\text{b}}^{\text{red}} $ and $m_{\ell\text{b}}^{\text{reco}}\ (\text{P}_{\text{gof}}<0.2)$, each used to constrain specific related systematics.
Observable $m_{\text{t}}^{\text{fit}}$ is treated as parameterized distribution, whereas other parameters are treated as binned using eight bins per observable.
The effect of adding each observable in the specified order on the expected total uncertainty is presented in Fig.~\ref{fig:ljets}.

The leading impacts and nuisance pulls are also presented in Fig.~\ref{fig:ljets}.
The largest uncertainty sources are related to the b-jet energy scale and final state radiation (FSR).
A significant pull is observed in $\text{q}\rightarrow \text{qg}$ FSR scale, which is related to a shift in $m_\text{W}^{\text{reco}}$ peak.
This is the most precise individual measurement to date, yielding a result of
$m_{\text{t}} = 171.77 \pm 0.37$ GeV.
The statistical uncertainty in data is estimated to be 0.04 GeV.\\

\noindent
\textbf{CMS boosted top}\\
A top mass measurement using boosted top quarks by CMS with the full Run 2 data of 138 fb$^{-1}$ is presented in Ref.~\cite{boosted}.
Top quark decay products are reconstructed as single jets using the XCone jet reconstruction algorithm \cite{xcone}, with a distance parameter of $\text{R}=1.2$, transverse momentum cut of $p_\text{T} > 400$ GeV and three subjets $N_{\text{sub}} = 3$.

The top quark mass is inferred from the differential cross section as a function of the XCone jet mass $m_{\text{jet}}$.
Analytical calculations in perturbative QCD are not possible for this topology at the moment and therefore the measurement is used to determine the top quark mass in \POWHEG \cite{powheg}.

A specific jet mass calibration is performed using the $m_\text{W}$ peak by applying two correction factors: $f^{\text{JEC}}$ for the subjet energy scale and $f^{\text{XCone}}$ for the XCone jet energy scale.
Also a calibration for the FSR is done based on the prior knowledge that $\alpha_\text{S}^{\text{FSR}}(m_\text{Z}^2)$ value in the CP5 tune is not optimal for $\text{t}\bar{\text{t}}$ jet substructure modeling.
The best fit value for $\alpha_\text{S}^{\text{FSR}}(m_\text{Z}^2)$ is obtained by fitting to data using the N-subjettiness ratio $\tau_{32} = \tau_3 / \tau_2$, which is sensitive to the angular distribution of the energy density inside jets.

\begin{figure}[H]
\centering
\includegraphics[width=0.49\textwidth, valign=c]{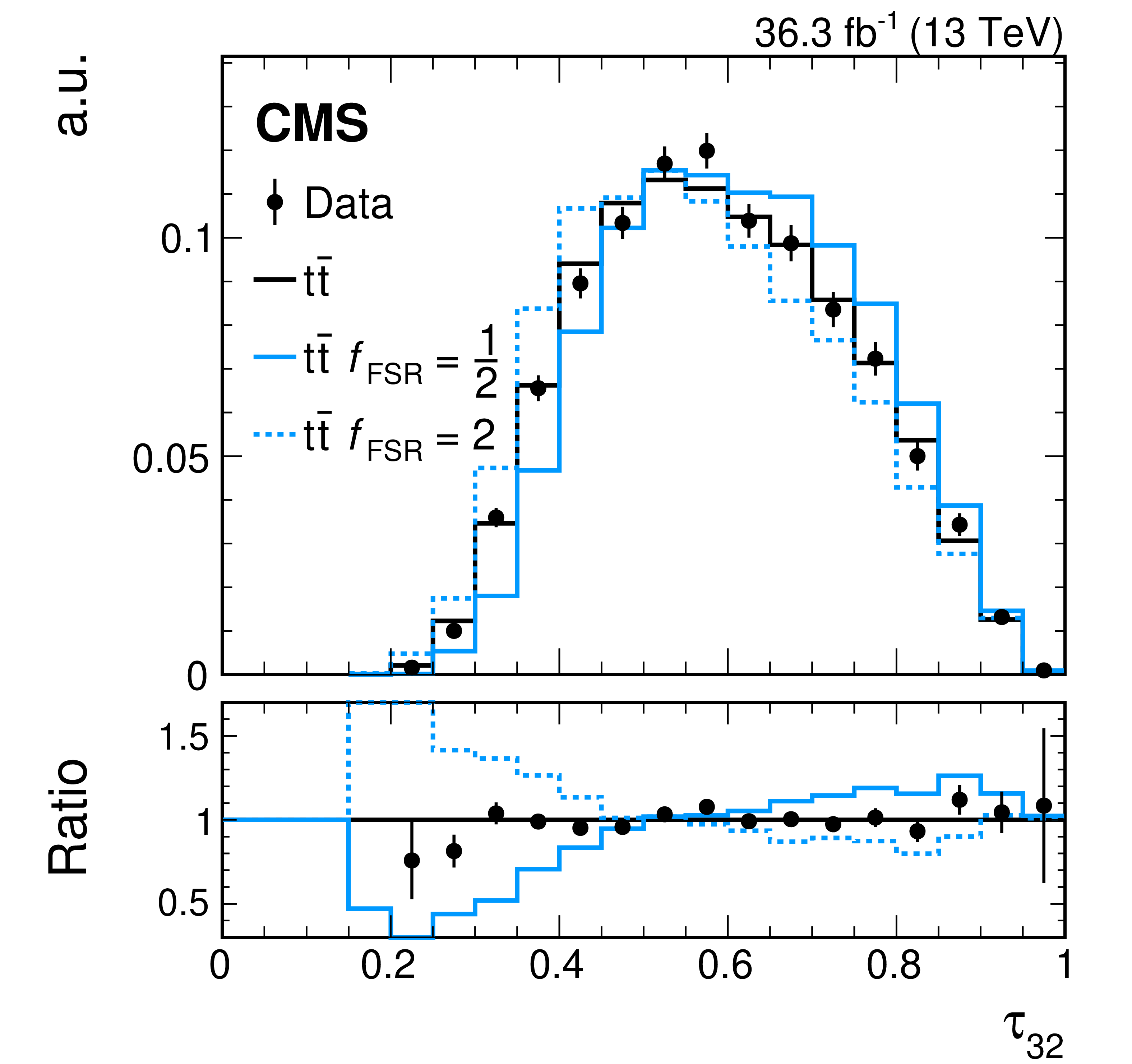}
\includegraphics[width=0.5\textwidth, valign=c]{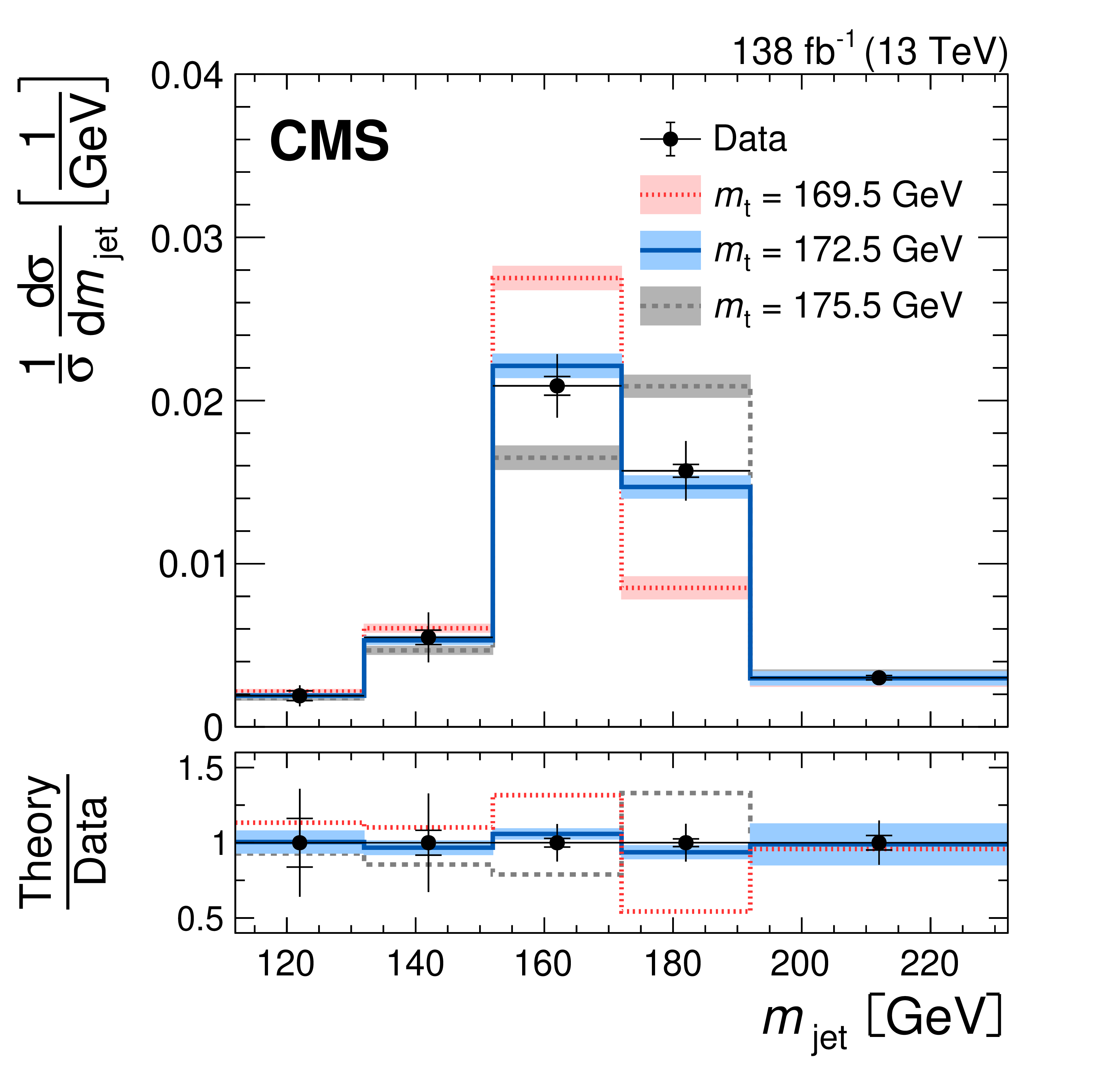}
\caption{Left: The normalised $\tau_{32}$ distribution with different FSR scales compared to data.
Right: The normalised differential cross section as a function of $m_{\text{jet}}$. \cite{boosted}}
\label{fig:normdiff}
\end{figure}

A $\chi^2$ fit for the normalised differential cross section presented in Fig.~\ref{fig:normdiff} is used to extract the top quark mass.
This measurement leads to a value of $m_{\text{t}} = 173.06 \pm 0.84$ GeV, which improves the precision by a factor larger than three compared to the previous result \cite{prevboosted}.

\noindent
\textbf{ATLAS soft muon tagging}\\
In the measurement in Ref.~\cite{softmuontagging} conducted by the ATLAS collaboration, a leptonic observable $m_{\ell\mu}$ is used in the $\text{t}\bar{\text{t}}$ lepton and jets channel on a dataset of 36 fb$^{-1}$. 
This observable, defined as the invariant mass of a lepton and a muon from the semileptonic decay of a b hadron originating from the same top quark, is designed to be less sensitive to jet energy scale related uncertainties, albeit with increased sensitivity to b-fragmentation modeling.

The jet associated to the b quark is identified using the displaced jet tagging and the soft muon tagging (SMT).
The SMT tight muon is required to be within a distance of $\Delta R <0.4$ from the selected jet candidate.
To ensure that the SMT muon originates from the same top quark as the lepton from the W boson decay, an additional requirement of $\Delta R ($SMT $\mu$, $\ell$)$< 2$ is enforced.

Due to the increased sensitivity to b-fragmentation modeling, a specific fit is performed for the parameter $r_\text{b}$ in the Lund-Bowler parameterization.
A binned $\chi^2$ test on the observable $x_\text{B} = 2p_\text{B} \cdot p_\text{Z}/m_\text{Z}^2$ is performed using data from the ALEPH, DELPHI and OPAL experiments at LEP, and SLD at SLC \cite{aleph, delphi, opal, sld}, leading to a value of $r_\text{b} = 1.05 \pm 0.02$.
Additionally, the production fractions and branching ratios for b and c hadrons are rescaled to world averages.

A binned-template profile likelihood fit is used to extract the $m_\text{t}$ in a range $m_{\ell\mu} = [15,80]$ GeV.
An additional uncertainty related to the gluon-recoil scheme is considered.
By default, in the parton shower of a b quark in the $\text{t}\rightarrow \text{Wb}$ system, a radiated gluon is set to recoil against the b quark \cite{recoil}.
Tests in which the gluon is set to recoil against the top quark were performed, suggesting an additional uncertainty of $\pm 0.25$ GeV that is added outside of the profile likelihood fit.
However, since there is no dedicated tune yet, this effect is likely to be overestimated.
As a result, this measurement yields a value of $m_\text{t} = 174.41 \pm 0.81$ GeV.
Leading uncertainties are related to the branching ratios of the b quark and the gluon-recoil effect.\\

\begin{figure}[H]
\centering
\includegraphics[width=0.475\textwidth, valign=c]{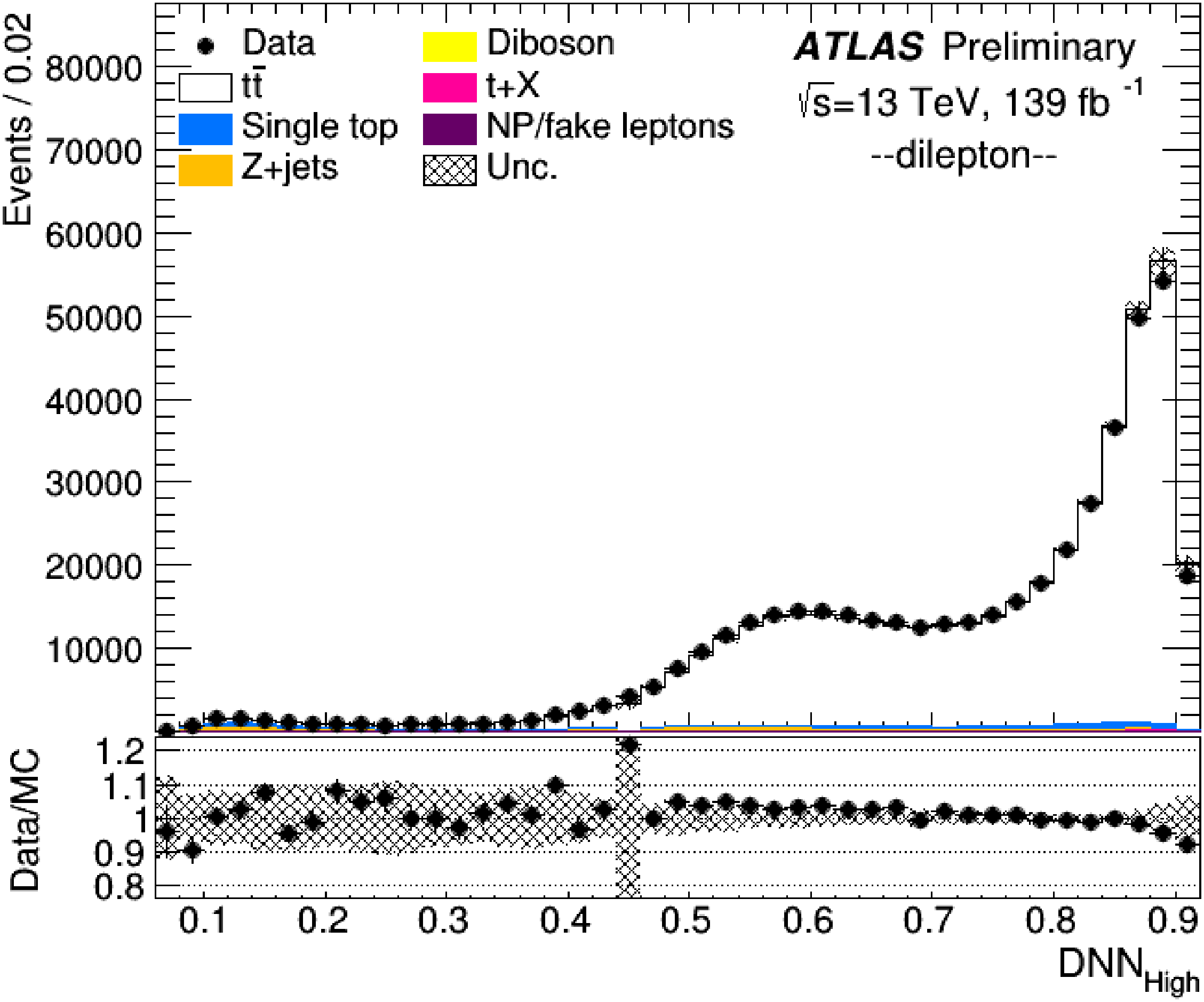}
\includegraphics[width=0.485\textwidth, valign=c]{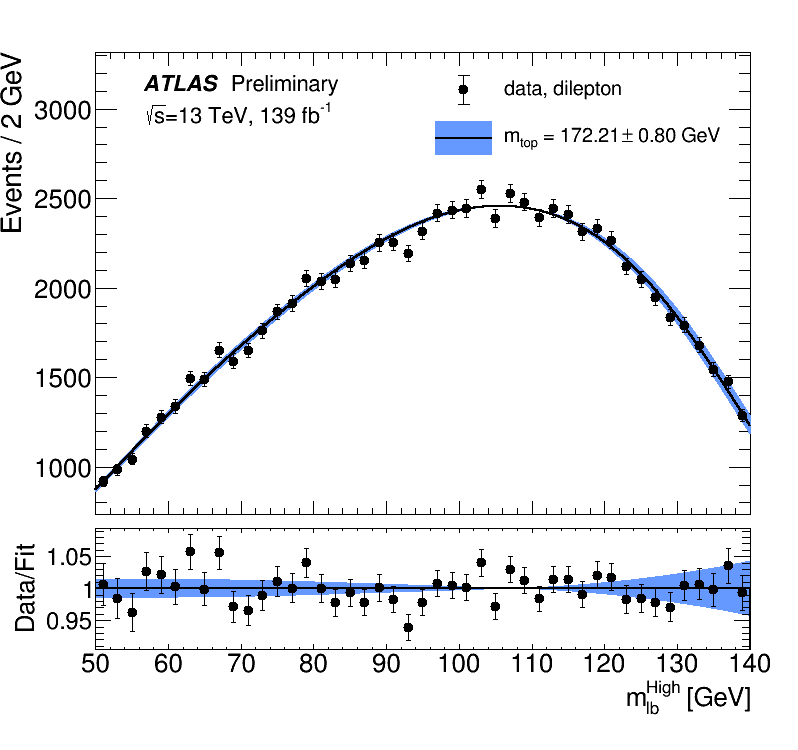}
\caption{Left: The $\text{DNN}_{\text{High}}$ distribution. 
Right: The fitted $m_{\ell\text{b}}^{\text{High}}$ distribution. \cite{ATLAS-CONF-2022-058}}
\label{fig:dnnhigh}
\end{figure}

\noindent
\textbf{ATLAS $\text{t}\bar{\text{t}}$ dilepton}\\
The full Run 2 statistics of 139 fb$^{-1}$ in fully leptonic decays of $\text{t}\bar{\text{t}}$ events are used in the measurement by the ATLAS collaboration in Ref.~\cite{ATLAS-CONF-2022-058}.
A pair of opposite sign charged leptons and two b-tagged jets are required in the event selection.
To increase the $\ell \text{b}$-pairing efficiency, a deep neural network (DNN) is utilized.
The DNN takes different kinematic variables related to leptons, b-tagged jets and
$\ell$b-pairs as an input.
The DNN output score corresponding to the best event hypothesis is referred to as $\text{DNN}_{\text{High}}$ and its distribution is shown in Fig.~\ref{fig:dnnhigh}.
The purity is increased by applying a cut $\text{DNN}_{\text{High}} > 0.65$.
A signal reweighting based on next-to-next-to-leading-order calculations is performed, resulting in a mass shift of $\Delta m_\text{t} = +0.1 \pm 0.1$ GeV, which is covered by the scale-variation uncertainties.
Off-shell and non-resonant effects have been studied using the bb4l generator in \POWHEG resulting in a shift of $\Delta m_\text{t} = -0.28 \pm 0.13$ GeV, which is not considered since it is covered by the modeling uncertainties.

An unbinned ML fit is performed in a range $50 \text{ GeV } < m_{\ell\text{b}}^{\text{High}} < 140 \text{ GeV}$ to extract the top quark mass.
Data and the template fit of the $m_{\ell\text{b}}^{\text{High}}$ distribution are presented in Fig.~\ref{fig:dnnhigh}.
The gluon-recoil uncertainty is estimated by the full difference between fits to pseudo-data where the recoil-to-top setting is switched on, yielding an uncertainty of $\pm 0.39$ GeV.
The resulting top mass value is $m_\text{t} = 172.21 \pm 0.80$ GeV, which improves the 8 TeV measurement by $17~\%$, if the gluon-recoil effect is not taken into account.\\

\noindent
\textbf{CMS $\text{t}\bar{\text{t}}$+1j pole mass}\\
The top quark pole mass is extracted using a normalised differential cross section measured at the detector level and unfolded using the ML method with profiled nuisance parameters in the analysis by the CMS collaboration in Ref.~\cite{polemass} on a data set of $36.3$ fb$^{-1}$.
The measurement is performed as a function of the variable $\rho = 2m_0 / m_{\text{t}\bar{\text{t}}+\text{jet}}$, where the scaling constant $m_0 = 170$ GeV.
The measurement uses a fully leptonic channel with one external jet.

Two kinematic reconstruction methods ("full" and "loose") are used.
To enhance the precision of the reconstruction of the observable $\rho$, a multivariate analysis method based on a regression neural network is used, where the target variable is the parton-level $\rho$.
From approximately one hundred initial input parameters, the ten most relevant are selected.
The same architecture is utilized for event classification into three output classes: $\text{t}\bar{\text{t}}$+jet, Z+jets and $\text{t}\bar{\text{t}}$+0jets.

The resulting $m_\text{t}^{\text{pole}}$ value is obtained from a $\chi^2$ fit of the normalised differential cross section at next-to-leading-order, without making any assumptions about the relationship between $m_\text{t}^{\text{MC}}$ and $m_\text{t}^{\text{pole}}$.
Using two parton distribution functions, the following results are obtained: ABMP16NLO yields $m_\text{t}^{\text{pole}} = 172.93 \pm 1.36$ GeV, and CT18NLO yields $m_\text{t}^{\text{pole}} = 172.13 \pm 1.43$ GeV. 
Both results are in good agreement with the previous $\text{t}\bar{\text{t}}$+jet measurement performed by the ATLAS collaboration \cite{ttjetatlas}.

\section{Conclusion}
Recent top quark mass measurements by the ATLAS and CMS collaborations were reviewed. 
As the direct top quark mass measurements become increasingly precise, a comprehensive understanding of  the underlying systematic uncertainties, such as those related to FSR and gluon-recoil effects, becomes crucial. 
The definition of the top quark mass itself introduces additional uncertainties in the direct measurements. 
Indirect measurement techniques provide an alternative, offering a theoretically clearer mass definition, such as in the $m_\text{t}^{\text{pole}}$ measurements.
As the experimental methods evolve and the precision improves, a collaborative effort between theorists and experimentalists is essential to advance the field of top quark mass measurements.






\end{document}

%% file: econfmacros.tex
\def\beq{\begin{equation}}
\def\eeq#1{\label{#1}\end{equation}}
\def\eeqn{\end{equation}}

\def\beqa{\begin{eqnarray}}
\def\eeqa#1{\label{#1}\end{eqnarray}}
\def\eeqan{\end{eqnarray}}

\let\bar=\overbar

\def\Dslash{\not{\hbox{\kern-4pt $D$}}}
\def\dslash{\not{\hbox{\kern-2pt $\del$}}}

\def\msb{{\bar{\ssstyle M \kern -1pt S}}}

